\documentclass[aps, pra, reprint, longbibliography,
superscriptaddress]{revtex4-1}

\usepackage{hyperref}
\usepackage{graphicx}
\usepackage[utf8]{inputenc}
\usepackage{xcolor}
\usepackage{amsmath, amssymb}
\usepackage{physics}
\usepackage{ulem}

\begin{document}

\preprint{APS/123-QED}

\title{Creating and melting a supersolid by heating a quantum dipolar system}

\author{R. Bombín}
\email{raul.bombin@u-bordeaux.fr}
\affiliation{Université de Bordeaux, 351 Cours de la Libération, 33405 Talence, France}
\author{J. Boronat}
\affiliation{Departament de F\'isica, Universitat Polit\`ecnica de Catalunya, Campus Nord B4-B5, 08034 Barcelona, Spain}
\author{F. Mazzanti}
\affiliation{Departament de F\'isica, Universitat Polit\`ecnica de Catalunya, Campus Nord B4-B5, 08034 Barcelona, Spain}
\author{J. S\'anchez-Baena}
\email{juan.sanchez.baena@upc.edu}
\affiliation{Departament de F\'isica, Universitat Polit\`ecnica de Catalunya, Campus Nord B4-B5, 08034 Barcelona, Spain}

\date{\today}

\begin{abstract}

Recent experiments have shown that rising the temperature of a dipolar gas under certain conditions leads to a
transition to a supersolid state.
Here, we employ the path integral Monte Carlo method, which exactly accounts
for both thermal and correlation effects, to study that phenomenology in a system of $^{162}$Dy atoms in the canonical ensemble.
Our microscopic description allows to quantitatively characterize the emergence of spatial order and superfluidity,
the two ingredients that define a supersolid state. Our calculations prove that temperature on its own can promote the formation of a supersolid in a dipolar system. Furthermore, we bridge this exotic phenomenology with the more usual melting of the supersolid at a higher temperature. Our results offer insight into the interplay between thermal excitations,
the dipole-dipole interaction, quantum statistics and supersolidity.

\end{abstract}

\maketitle

Supersolidity corresponds to the counter-intuitive state of matter that
combines global phase coherence and crystalline order. Conceived more than fifty years ago~\cite{Andreev:JETP:1969,Chester:PRA:1970,Leggett:PRL:1970},
early experimental efforts to observe supersolidity focused on Helium-based
systems~\cite{Chan2013}. However, despite the apparent initial success in the
detection of a tiny superfluid fraction~\cite{Kim2004,Kim2004:Science}, these
experimental observations were later attributed to the change of the elastic properties
of the solid, rather than to the presence of a superfluid flow.

Years later, thanks to the rapid progress in the
trapping and control of ultracold atoms other platforms have revealed themselves
as good candidates to observe supersolidity. Among these, dipolar quantum
gases have become the preferred option for this purpose. This is due to the
peculiar combination of traits of the dipole-dipole interaction, mainly its
long-range character and its anisotropy.
Even if these systems were predicted
to collapse at the mean-field level~\cite{Lahaye:PRL:2008,Pfau:NatPhys:2008},
beyond mean field calculations and, remarkably experiments, showed that
quantum fluctuations crucially arrest the collapse~\cite{Lima:2011eq,Lima:2012hr},
giving rise to the formation of dipolar droplets~\cite{kadau16,Chomaz:2016do}.

More precisely, it has been the study of arrays of such droplets what has given
access to supersolid states~\cite{Natale:PRL:2019,
Tanzi:Nature:2019,Modugno:PRL:2019,Tanzi:Science:2021,HertkornPRX21,
Ferlaino:PRL:2021,norcia21,Biagioni:PRX:2022}.
In this case,
the dipolar supersolid consists of clusters made up of hundreds or thousands of
dipolar atoms surrounded by a low density cloud that provides the
necessary phase coherence~\cite{Chomaz_2023}. This is rather different
to the hypothetical supersolid state in commensurate hcp $^4$He.

\begin{figure*}[t]
\centering
\includegraphics[width=\linewidth]{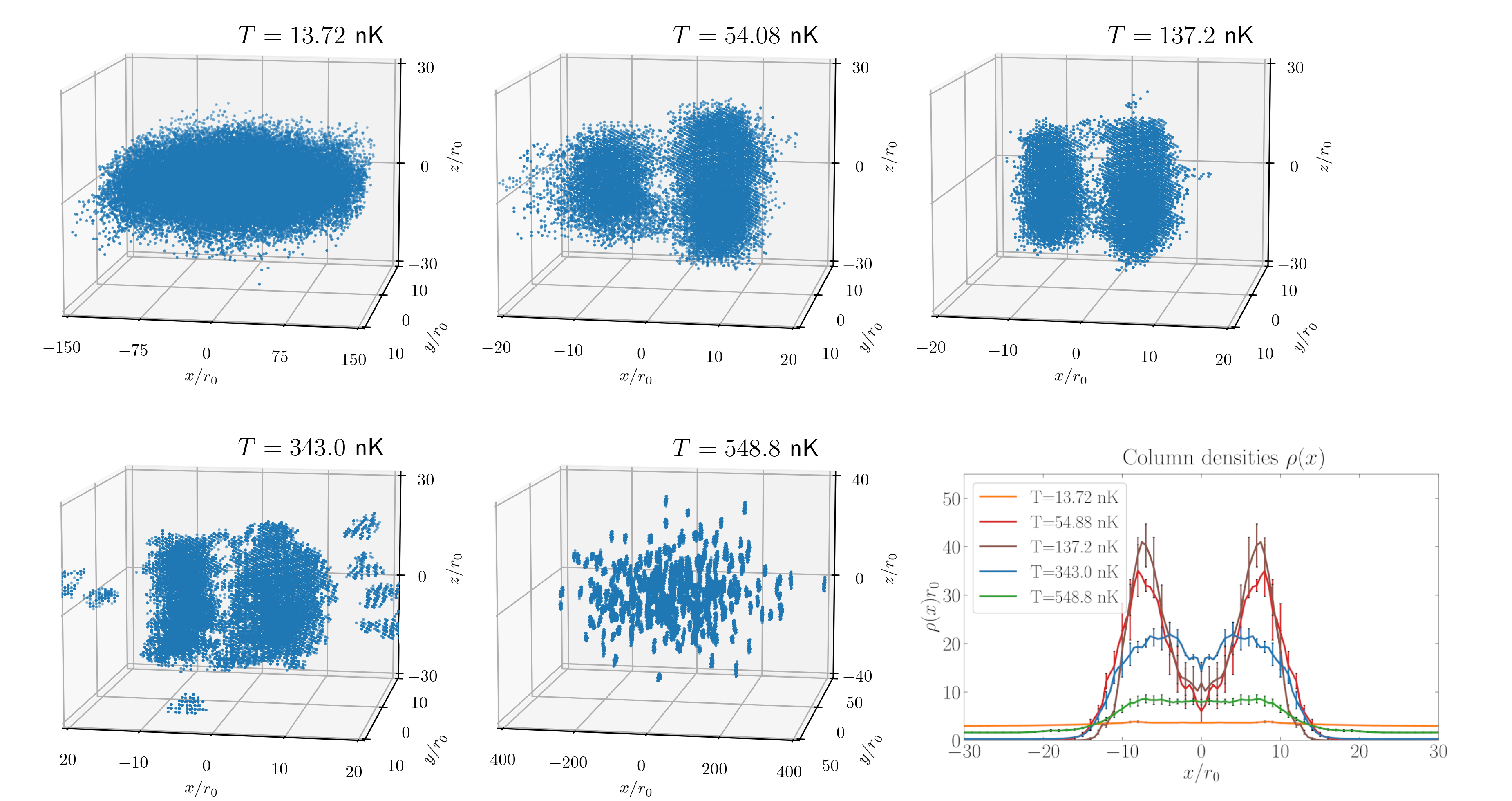}
\caption{Snapshots corresponding to the formation and melting of
 supersolid states with increasing temperature, obtained
 from the PIMC simulations. The plot in the bottom right displays the
 corresponding symmetrized column densities along the $x$-axis, which have
 been computed by averaging several realizations at fixed temperature for a
 scattering length value $a_s = 70 a_0$. }
\label{fig_1}
\end{figure*}

The effect of temperature in the supersolid properties of dipolar systems is a
thorough and relatively unexplored subject. Recently, pioneering experimental
and
theoretical results have shown that, under certain conditions, thermal
fluctuations can promote supersolidity in these systems~\cite{Ferlaino:PRL:2021,baena22}. In the experiment of
Ref.~\cite{Ferlaino:PRL:2021}, the state of a dipolar gas is compared at different
stages of an evaporative cooling sequence where the gas hosts the same number
of condensed atoms, but different total atom number and temperature.
There, it was shown that within the nano-Kelvin range, the hotter state
features a density modulation that is absent in the colder one. In other words,
a solid like structure appears after heating the system.
Nevertheless, since the total particle number influences the promotion of spatial order in dipolar systems, these experimental results do not allow to disentangle the role played by particle number and temperature in the rise of the supersolid. These results were initially rationalized by including perturbative thermal
corrections in Bogoliubov theory~\cite{baena22}.
However this approach is limited to the study of low temperature excitations and does not
allow to quantitatively estimate the superfluid fraction of the system. The latter, together with the emergence of
spatial order, are the key properties to properly characterize a supersolid state.

In this Letter, we provide insight into the different physical mechanisms that
collaborate and compete to give rise
to such an exotic phenomenology, namely quantum delocalization, dipolar
interaction, quantum statistics, and thermal effects. Moreover, since we work in the canonical ensemble, we are able to assess the role played by temperature alone in the formation and melting of a supersolid.
To this purpose, we perform ab initio path integral Monte Carlo
(PIMC)~\cite{Ceperley_1995}
calculations that take into account both quantum correlations and thermal fluctuations to all orders. In our implementation we use
a fourth order Chin's action~\cite{Chin2002} with 180 intermediate coordinates (beads) that provides converged results for the lowest temperatures considered.
We make use of the worm algorithm to sample
efficiently permutations between identical particles~\cite{Boninsegni2006}.
In this way, we can estimate the superfluid
fraction and thus, quantitatively check whether the thermally modulated
states are indeed supersolid. Furthermore, the inclusion, or not, of
permutations in our calculations allows us to disentangle the role played by
quantum statistics from other quantum effects such as delocalization, which is
an unexplored question. Additionally, the PIMC algorithm is suitable to study a
vast range of temperatures, thus filling the gap between the ultracold regime
and larger temperatures where superfluidity vanishes and a solid is
expected to melt.

We consider a trapped system of
$N=400$ $^{162}$Dy dipolar atoms with  trapping frequencies $
(\omega_x,\omega_y,\omega_z) = 2 \pi \times (228, 3649, 3649)$ Hz. This results
into a mean density of order $n \sim  7 \times 10^{14}$ cm$^{-3}$, a value that
matches the experimental data~\cite{hertkorn21} corresponding to $N=40000$ and
$(\omega_x,\omega_y,\omega_z) = 2 \pi \times (30, 89, 108)$ Hz.
The inter-particle interaction is modeled as
\begin{equation}
 V({\bf r}) = \frac{C_{12}}{r^{12}} - \frac{C_6}{r^6} +
 \frac{C_{\rm dd}}{4 \pi} \frac{ \left( 1 - 3 \cos^2 \theta  \right) }{r^3}
 \ ,
 \label{potential}
\end{equation}
corresponding to a fully polarized system along the $z$-axis, with
${\bf r} = {\bf r}_i - {\bf r}_j$, $r = \abs{{\bf r}}$, and $\theta$ the angle between the vector ${\bf r}$ and the $z$ axis.
In the following we use reduced units, with the
characteristic length and energy scales defined by
$r_0 = m C_{\rm dd}/(4\pi \hbar^2)$ and $\epsilon = \hbar^2/(m r_0^2)$. In
these units, $C_6 = 0.02978$ for $^{162}$Dy atoms~\cite{Li2017}, while
the value of $C_{12}$ is obtained by fixing the $s$-wave
scattering length $a_s$ of the interaction to the
desired value through a T-matrix calculation~\cite{bottcher19}.

\begin{figure}[b]
\centering
\includegraphics[width=\linewidth]{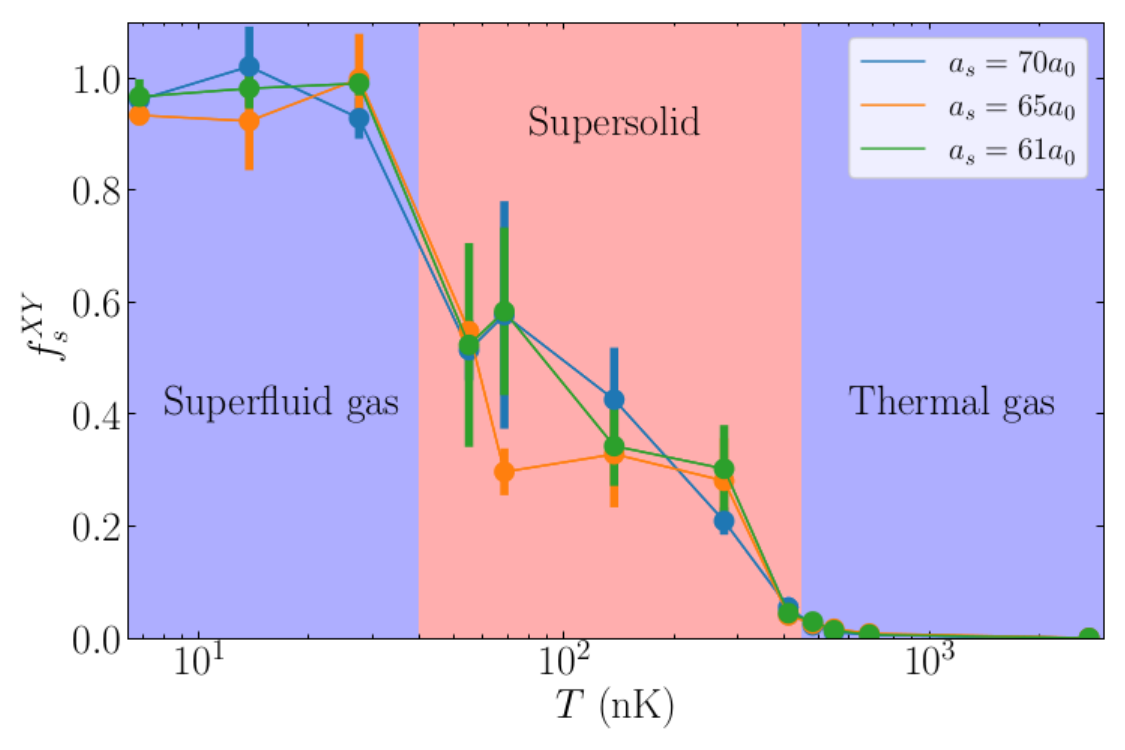}
\caption{ Superfluid fraction along the $x$-$y$ plane as a function of the
 temperature for three different values of the scattering length. The blue
 and red regions correspond to unmodulated and modulated equilibrium
 configurations, respectively. }
\label{fig_2}
\end{figure}

\begin{figure*}[t]
\centering
\includegraphics[width=0.9\linewidth]{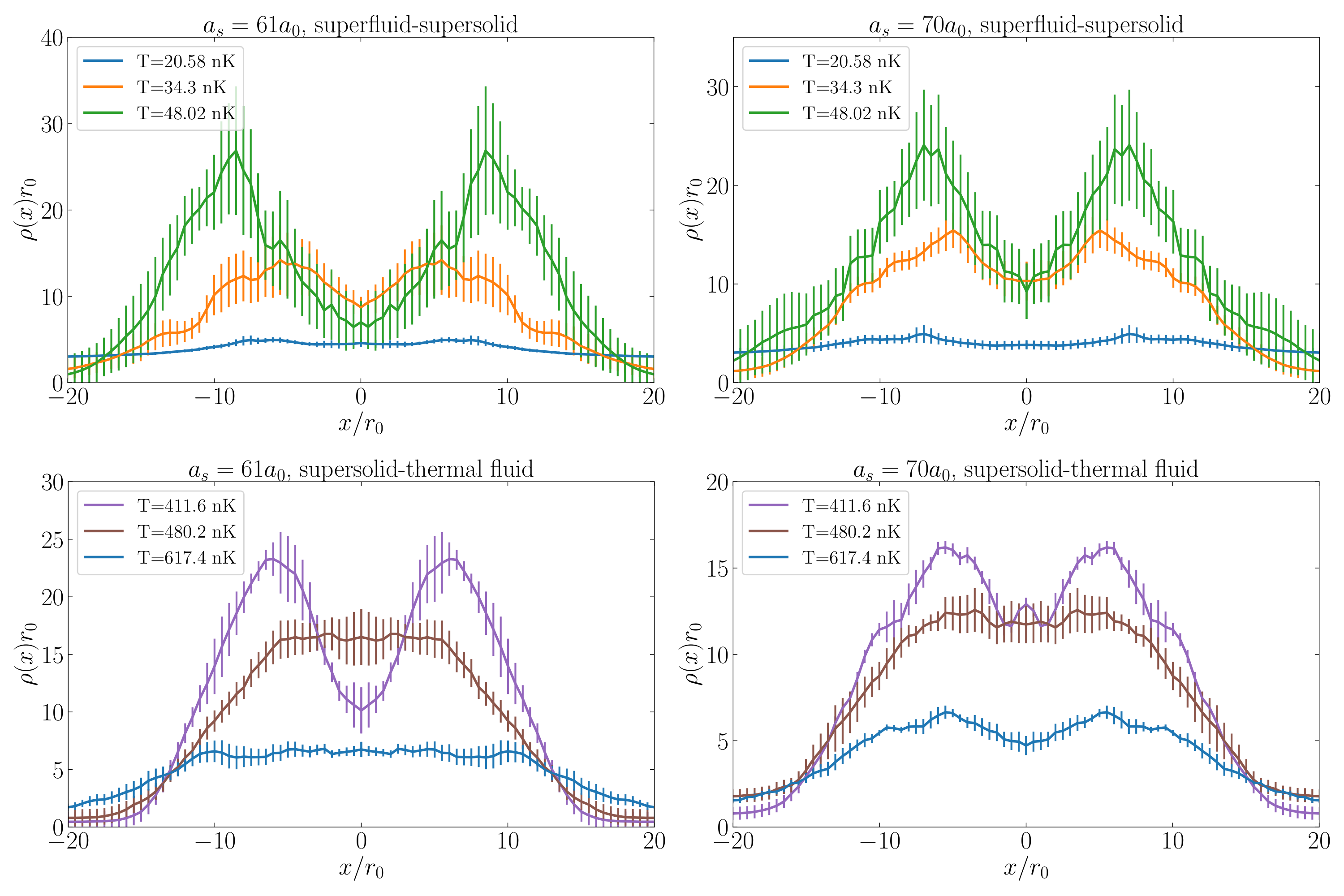}
\caption{ Emergence of the modulations (top) and melting (bottom) for two
 values of the scattering length: $a_s = 61 a_0$ (left) and $a_s = 70 a_0$
 (right). The column densities are symmetrized around the origin.
 The plots show the result of averaging several realizations yielding
 two-droplets. }
\label{fig_3}
\end{figure*}

As it is known, tuning the scattering length
of the system at zero temperature through the $C_{12}$ parameter
allows to cross a transition from an unmodulated gas to a
supersolid, and vice-versa.
For $^{162}$Dy atoms in the trap considered here, it was shown
that the transition from a BEC gas to a supersolid state occurs for $a_s
\le$~60~$a_0$, with $a_0$ the Bohr radius~\cite{Bombin2024}.
Our goal is to study the possible thermal transition from a gaseous BEC to a
supersolid at finite temperature.
With this purpose, we consider three different scattering lengths,
$a_s/a_0 =61$, $65$, and $70$, which yield an unmodulated ground state at $T = 0$.
Our calculations reveal that for the lowest temperature considered ($T = $~13.72 nK)
an unmodulated BEC appears for the three values of the scattering length,
in agreement with the zero temperature path integral ground state calculations
of Ref.~\cite{Bombin2024}.
This is depicted in Fig.~\ref{fig_1} for $a_\textrm{s} = $~70~$a_0$.
Interestingly, as temperature rises density modulations appear in the range $\left[40,450\right]$~nK.
If temperature is further increased the solid structure is melted, which is the commonly expected physical scenario.
It is important to notice that for intermediate temperatures,
different PIMC realizations starting with different initial conditions
give rise to states with one or two droplets, meaning that these configurations are very close in free energy.
Because of this, and due to the existence of free energy barriers between these states, one expects these different configurations to be accessible
in an experimental realization.
Indeed, the phenomenology depicted in Fig.~\ref{fig_1} reminds the
one observed in Ref.~\cite{Ferlaino:PRL:2021}. However,
the experiment was performed for a fixed number of atoms in the condensate while
we work in the canonical ensemble, with a total number
of atoms constant.
This implies that the surprising emergence of a supersolid by increasing the temperature is not a feature dependent on the specific conditions of the experiment of Ref.~\cite{Ferlaino:PRL:2021}, but rather a general and robust phenomenon in dipolar systems.

\begin{figure*}[t]
\centering
\includegraphics[width=\linewidth]{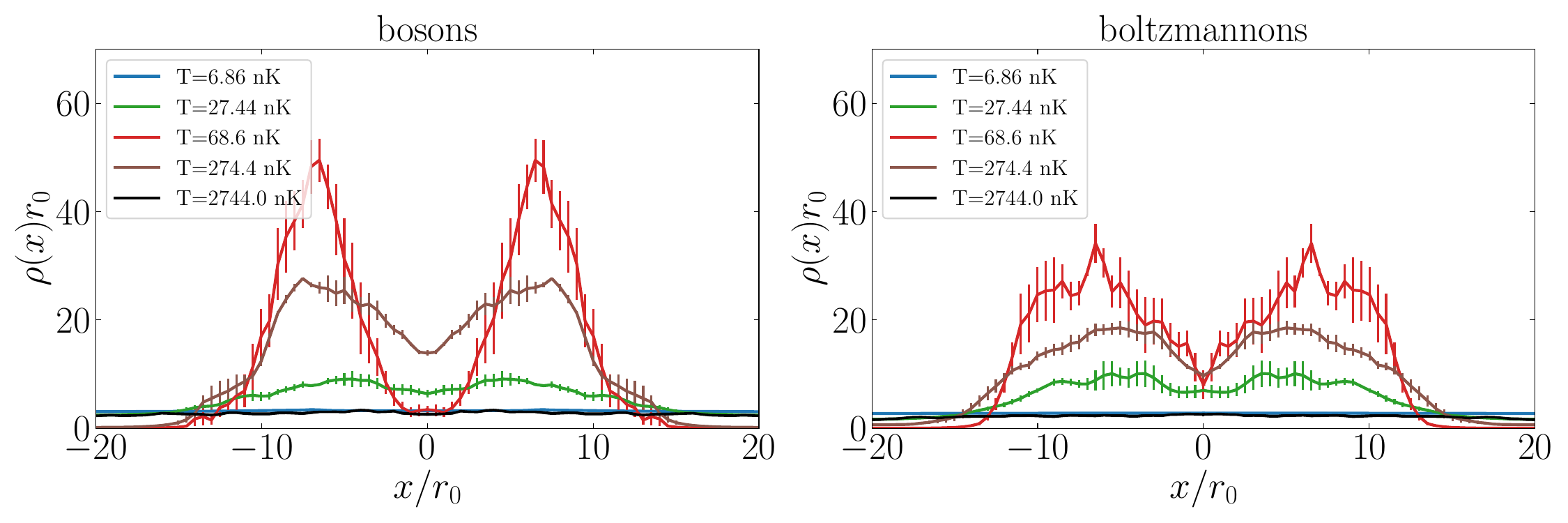}
\caption{ Symmetrized column densities at different temperatures for the
 system assuming bosonic (left) and Maxwell-Boltzmann (right) statistics, for
 a scattering length $a_s = 70 a_0$. The distributions are obtained from
 averages of several PIMC simulations producing two droplets as the lowest
 free-energy state. }
\label{fig_4}
\end{figure*}

Also in Fig.~\ref{fig_1}, the corresponding $x$-column densities, computed as
$n(x)=\int dx dy\, n(x,y,z)$, are shown.
It is important to remark that not only the BEC becomes modulated, but crucially the column
density profile becomes much tighter along the $x$-axis, which is compatible
with a significant increase of the density caused by temperature, consistent with Bogoliubov theory~\cite{baena22}.
Notably, although not shown in the figure,
all three values of $a$ display the emergence of modulations as temperature
increases, which is a clear signature of the robustness of the mechanism behind
this transition.
We have checked that the gas parameter is, in all cases, lower
than $10^{-2}$, meaning  that the system still lies close to the dilute regime.

In order to characterize the supersolid nature of the modulated states
appearing at intermediate temperatures, one needs to evaluate the corresponding
superfluid fraction.
The PIMC method allows for an exact estimation of the superfluid
properties of the system via the area estimator~\cite{Ceperley_1995}.
We evaluate the
ratio between the moment of inertia of the atomic cloud, when the
system is rotated around the $z$-axis $I^{\text{XY}}$, and the corresponding
classical value $I_c^{\text{XY}}$.
From this ratio, we obtain the superfluid
fraction in the $x$-$y$ plane
via the relation~\cite{Tanzi:Science:2021}
\begin{equation}
 f_s^{\text{XY}} = \frac{1 - I^{\text{XY}}/I_c^{\text{XY}} }{1 - \beta^2} \ ,
\end{equation}
where $\beta = \frac{\langle x^2 - y^2 \rangle}{\langle x^2 + y^2 \rangle}$ is
a geometric factor to account for the anisotropy of the cloud.
Notice that the squared averages
entering
in the definition of $\beta$ depend on each realization.
Figure~\ref{fig_2} shows $f_s^{XY}$ as a function of the temperature for the three
scattering length values considered in this work.
At low temperature, the system is fully superfluid, and does not show
modulations, therefore constituting a superfluid gas. As temperature starts to
rise above 40~nK, modulations arise and the superfluid signal is reduced,
although it is clearly distinct from zero. This
reduction of superfluidity is caused by both thermal effects and the appearance
of spatial structure (\textbf{c.f.} Fig.~\ref{fig_1}).
These results thus confirm that
heating the systems is not only
promoting spatial order, but also inducing a transition to a supersolid state. Finally, further heating the system causes the superfluid signal to vanish for temperatures approximately larger than 450~nK.

Combining the results of Figs.~\ref{fig_2} and~\ref{fig_3}, we can see that the superfluid signal and spatial structure disappear at very similar temperatures. At $T = 411$~nK we obtain a modulated state with a very small superfluid fraction $f_s^{XY} < 0.05$, which melts before reaching $T = 480$~nK. Thus,
a non superfluid solid is likely to appear only in a very narrow regime of temperatures.
This phenomenology contrasts with that of a bulk two-dimensional dipolar
system, where the supersolid emerges as a stripe phase.
In that case, it has been shown that the critical temperatures
for the two transitions (loss of modulation and loss of superfluidity) differ at least by a factor of five, with superfluidity being lost at a lower temperature~\cite{bombin19}.
The different thermal behavior of the two systems probably
lies on the role played by fluctuations.
Due to the balance between the repulsive and attractive parts of the dipole-dipole interaction, quantum and thermal fluctuations play a huge role at low temperatures in three dimensions~\cite{baena22}. However, in the 2D regime considered in Ref.~\cite{bombin19} the dipole-dipole interaction is always repulsive instead.
This reveals that the rich supersolid physics in dipolar systems may strongly
depend on the the interplay between geometry, interaction, and quantum
correlations.

The results of Fig.~\ref{fig_2} suggest that the superfluid-supersolid and supersolid-thermal gas transitions do not depend strongly
on the scattering length. To illustrate this, in Fig.~\ref{fig_3} we focus on the temperature ranges
$[20,50]$~nK and $[400,620]$~nK to study both thermal transitions at $a_s/a_0 =
61$, $70$.
The emergence and melting of
the supersolid states occurs at similar temperatures in both cases, meaning
that the formation of the supersolid is less sensitive to $a_s$ than in previous
calculations for a dipolar system in a tubular geometry under the
Bogoliubov approximation~\cite{baena22}.
Nevertheless, the density contrast is larger for the smallest scattering
length value.

Finally, it is interesting to address the question of the role played by
quantum statistics on this observed behavior.
It is known that weakly interacting bosonic systems below the critical
temperature manifest a bunching effect
as temperature is raised
(see for example, Ref.~\cite{Pascual2023} for its characterization in Bose-Bose mixtures). This effect can modify the density distribution of the system, as it happens in the case of Bose-Bose mixtures.
One may thus wonder whether the bunching effect is playing a role in the emergence of
supersolid states in our system.
In PIMC, it is possible to perform calculations replacing the dipolar bosons
by distinguishable quantum particles that follow Maxwell-Boltzmann
statistics, or boltzmannons. Technically, this is easily achieved by
not sampling permutation cycles in the PIMC algorithm.
We compare, in Fig.~\ref{fig_4}, the column density profiles along the $x$
axis for bosons and boltzmannons at different temperatures for a scattering
length $a_s=70 a_0$. Increasing the temperature in the boltzmannon
system leads also to the emergence of a modulated state.
This reveals that the bunching effect, if present, is not the
only driving mechanism behind the emerging of supersolid states after heating.
Notice however that the density modulations
are higher in the bosonic case, which implies that, while not being the source
for the emergence of supersolidity, the bosonic bunching effect enhances it.

In summary, we have studied the thermal evolution of a
trapped system of $N=400$ $^{162}$Dy atoms at finite temperature
by means of path integral Monte Carlo calculations.
Similarly to the phenomenology observed in the experimental realization of
Ref.~\cite{Ferlaino:PRL:2021,baena22} we observe that
rising the temperature of gaseous BEC in the nano-Kelvin regime leads to the emergence of a supersolid state, which we quantitatively characterize by exactly computing its density distribution and superfluid fraction. However, since our calculations are performed in the canonical ensemble, we are able to disentangle the effect of particle number and temperature in the rise of supersolidity, and show that temperature alone is sufficient to promote a supersolid state.
Furthermore, we also characterize the subsequent supersolid melting by heating, and show that, for our parameters of choice, the temperatures at which superfluidity and spatial modulation are lost are remarkably close. This feature is clearly distinct from that of an infinite two-dimensional dipolar system~\cite{bombin19}, and stems from the different role played by quantum fluctuations. Finally, we have also addressed the role of quantum statistics in the emergence of supersolidity by heating. We show
that the emergence of spatial order by increasing the temperature is still present even if one
disregards the bosonic statistics. All in all, our results are not
accessible by perturbative methods and fill a gap
in the current theoretical knowledge
of ultracold dipolar systems, offering novel insight into the interplay between supersolidity and temperature.

For further work, deeper insight could be gathered by calculating the excitation spectra of the trapped dipolar system for the parameters considered in this work, as the energies of the excitations and their sensitivity to the
scattering length are crucial to understand the thermal effects. On a different note, an interesting extension of our work is the study of
a quasi-1D system of dipoles
in an infinitely extended tube. This will allow for the calculation of a
phase diagram in the thermodynamic limit, and the subsequent study of the
order of the phase transitions between the different phases, including the
high temperature ones inaccessible through Bogoliubov theory. Eventually, the
phase diagram could host a critical point, where all the expected phases
(superfluid, supersolid, normal solid and normal fluid) converge.
Finally, it is also interesting to consider the possible effects that
finite temperature may have in systems of dipolar molecules~\cite{Bigagli2024},
where the large tunability of interactions allows to access highly correlated
regimes, change the sign of the DDI or even its
anisotropy~\cite{karman2025:arxiv,deng2025:arxiv}.

We thank Thomas Pohl for useful discussions. We acknowledge financial support from Ministerio de Ciencia e
Innovaci\'on
MCIN/AEI/10.13039/501100011033
(Spain) under Grant No. PID2023-147469NB-C21  and
from AGAUR-Generalitat de Catalunya Grant No. 2021-SGR-01411.
R.B. acknowledges funding from ADAGIO (Advanced Manufacturing Research Fellowship Programme in the Basque – New Aquitaine Region) from the  European Union’s Horizon 2020 research and innovation programme under the Marie Sklodowska  Curie cofund Grant Agreement No. 101034379.

\bibliography{paper_temperature_dipoles}

\end{document}